\documentclass[twocolumn]{revtex4}
\usepackage{graphicx}

\begin{document}

\title{Spatial coherence singularities of a vortex field in nonlinear media}

\author{Kristian Motzek and Yuri Kivshar}
\affiliation{Nonlinear Physics Center, Research School of Physical
Sciences and Engineering, Australian National University, Canberra
ACT 0200, Australia}

\author{Grover A. Swartzlander, Jr.}
\affiliation{Optical Sciences Center, University of Arizona,
Tucson, Arizona 85721, USA}

\begin{abstract}
We study the spatial coherence properties of optical vortices
generated by partially incoherent light in self-focusing nonlinear
media. We reveal the existence of phase singularities in the
spatial coherence function of a vortex field that can characterize
the stable propagation of vortices through nonlinear media.
\end{abstract}

\maketitle

Vortices are known to occur in coherent systems having a zero
intensity at the center and a well-defined phase associated with
the circulation of momentum around the helix axis~\cite{berry}.
The last decade has seen a resurgence of interest in the study of
{\em optical vortices}~\cite{list}, owing in part to readily
available computer-generated holographic techniques for creating
vortices in laser beams.

If a vortex-carrying beam is partially incoherent, the phase front
topology is not well defined, and statistics are required to
quantify the phase. In the incoherent limit neither the helical
phase nor the characteristic zero intensity at the vortex center
can be observed. However, several recent studies have shed light
on the question how phase singularities can be unveiled in
incoherent light fields propagating in linear media
~\cite{SchoutenOL03,PalaciosPRL04}. In particular, Palacios {\em
et al.}~\cite{PalaciosPRL04} used both experimental and numerical
techniques to explore how a beam transmitted through a vortex
phase mask changes as the transverse coherence length at the input
of the mask varies. Assuming a quasi-monochromatic, statistically
stationary light source and ignoring temporal coherence effects,
they demonstrated that robust attributes of the vortex remain in
the beam, most prominently in the form of {\em a ring dislocation}
in the cross-correlation function.

Propagating in {\em nonlinear coherent systems}, optical vortices
become unstable when the nonlinear medium is self-focusing, and
this effect has been observed experimentally in different
nonlinear systems~\cite{book}. However, stable propagation of
optical vortices in self-focusing nonlinear media has been
recently demonstrated, both theoretically and experimentally, for
the vortices created by self-trapping of partially incoherent
light when the spatial incoherence of light exceeds a certain
threshold~\cite{JengPRL04}.

The purpose of this Letter is twofold. First, we study numerically
{\em the spatial coherence function} of a vortex beam propagating
in a self-focusing nonlinear medium and reveal its importance for
the study of singular beams in nonlinear media. Second, we provide
a deeper physical insight into the effect of vortex stabilization
by partially coherent light reported previously.

We consider a phase singularity carried by a single vortex beam.
To  simulate numerically the propagation of partially coherent
light through a nonlinear medium, we use the coherence density
approach~\cite{ChristodoulidesPRL97} based on the fact that an
incoherent light source can be thought of as a superposition of
(infinitely) many coherent components $E_j$ that are mutually
incoherent, having slightly different propagation directions:
$E(\mathbf{r},t)=\sum_j E_j(\mathbf{r}) \exp(i\mathbf{k}_{\perp j}
\mathbf{r})\exp(i\gamma_j(t))$, where $\mathbf{k}_{\perp j}=
k(\alpha_j\mathbf{e}_x+ \beta_j\mathbf{e}_y)$ is the transverse
wave vector of the $j$-th component, having direction cosines
$\alpha_j$ and $\beta_j$,
$\mathbf{r}=x\mathbf{e}_x+y\mathbf{e}_y$, $\gamma_j(t)$ is a
random variable varying on the time scale of the coherence time of
the light source, and $k=2\pi/\lambda$ is the wavenumber. The
vortex is introduced via a phase mask at the input face ($z=0$) of
the medium. To avoid complexities that may arise from incoherent
light sources having abrupt boundaries, we assume the source has a
Gaussian profile
\begin{eqnarray}
  \label{eq:1}
  E_j(\mathbf{r})=\left(\frac{1}{\sqrt{\pi}\theta_0}e^{-(\alpha_j^2+
      \beta_j^2)/\theta_0^2}\right)^{1/2} A(\mathbf{r}).
\end{eqnarray}
Here $\theta_0$ is a parameter that controls the beam's coherence,
$A(\mathbf{r})=(r/w_0)^2\exp(-r^2/\sigma^2)\exp(i\varphi)$ is the
vortex profile, and $\varphi$ is the angular variable.

Scaling the lengths in the transverse directions to $x_0=1\mu m$
and the length in propagation direction to $z_0=2kx_0^2$, where we
chose $k=2\pi/(230nm)$, the propagating field $E_j(\mathbf{r},z)$
can be described by the nonlinear Schr\"odinger equation:
\begin{eqnarray}
  \label{eq:2}
  i\frac{\partial  E_j(\mathbf{r},z)}{\partial z} +
  \nabla_\perp^2 E_j(\mathbf{r},z) + \eta(\mathbf{r},z)E_j(\mathbf{r},z) =0\, ,
\end{eqnarray}
where $\eta(\mathbf{r},z)$ accounts for the nonlinear refractive
index change in the material. We assume a photorefractive medium
with a saturable nonlinearity having a response time much longer
than the coherence time of the light source. In this case $\eta$
depends on the time-integrated intensity. The intensity is given
by $I=\sum_j|E_j|^2$. The nonlinear factor may thus be written in
the form,
$\eta(\mathbf{r},z)=I(\mathbf{r},z)/(1+sI(\mathbf{r},z))$, where
$s$ is a saturation parameter. Whereas numerical solutions of
Eq.~(\ref{eq:2}) may be readily computed using the coherence
density approach, we shall later adopt the equivalent multi-mode
theory~\cite{ChristodoulidesPRE01} to provide a physical basis for
our findings. The numerical simulations are performed using a
split-step method. In the examples shown below we use $N=1681$
components, $w_0=1.8$, $\sigma=1.5$ and $s=0.5$. We chose
$\theta_0=0.64^\circ$.

Second-order coherence properties of the propagating field may be
quantified by determining the the mutual coherence function
$\Gamma(\mathbf{r}_1,\mathbf{r}_2;z) =
\left<E^*(\mathbf{r}_2,z,t)E(\mathbf{r}_1,z,t)\right>$, where the
brackets represent an average over the net field
$E(\mathbf{r},z,t) = \sum_{j=1}^N E_j(\mathbf{r},z)
\exp(i\gamma_j(t))$. Again, we assume that the random phase
factors $\gamma_j(t)$ vary on a much faster timescale than the
response time of the medium. For the linear propagation, Palacios
{\em et al.}~\cite{PalaciosPRL04} demonstrated that the phase
singularities  occur in the cross-correlation
$\Gamma(-\mathbf{r},\mathbf{r})$ of an incoherent vortex, where
the origin of the coordinate system is chosen to coincide with the
vortex center.

\begin{figure}[htbp]
  \centerline{
    \includegraphics[width=50mm]{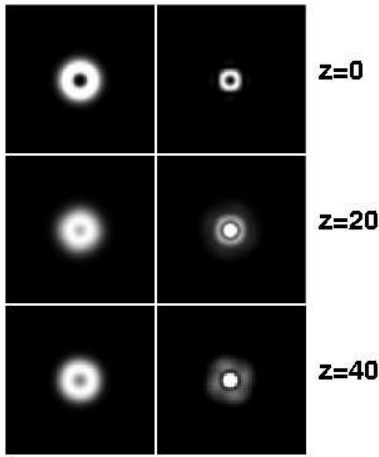}
  }
  \caption{
    Contour plots of the intensity (left column) and the modulus of the
    cross-correlation 
    (right column) of an incoherent vortex with $\theta_0=0.64^\circ$.
    Contrary to the case of the linear propagation, there is a local intensity
    minimum in the 
    beam's center. The cross-correlation, however, shows the same ring
    of phase singularities as predicted in the linear theory. The size
    is $35\times35\mu m$.
  }
  \label{fig1}
\end{figure}

In Fig.~\ref{fig1}, we show the results for the nonlinear
evolution of an incoherent vortex and the cross correlation.
First, we notice that the beam intensity has a local minimum in
the center of the vortex, even after propagating many diffraction
lengths. This is contrary to the case of the linear propagation
where a beam with the same degree of coherence $\theta_0$ has
maximum intensity in the center of the vortex after only a few
diffraction lengths. Also, if we had chosen to propagate an
incoherent ring of light without topological charge instead of an
incoherent vortex, we would also observe a maximum in the beam's
center. Thus we can state that the coherence function of the
vortex manifests itself in the intensity distribution of the light
beam after propagating through a nonlinear medium. In fact, the
intensity profile remains reminiscent of a vortex, even if the
intensity does not quite drop to zero in the center of the beam.

Looking at the beam's cross-correlation, we clearly observe a ring
of phase singularities in the cross-correlation
$\Gamma(-\mathbf{r},\mathbf{r})$. Thus, as the first result of our
numerical studies we state that the phase singularities predicted
for the incoherent vortices propagating in linear media also
survive the propagation through a nonlinear medium. This is not
self-evident, considering that in the nonlinear case the single
components that form an incoherent light beam do interact,
contrary to the linear case. A physically intuitive explanation
how this ring of phase singularities develops under linear
propagation is given in Ref.~[\onlinecite{PalaciosPRL04}].
However,  this issue becomes more complicated for the propagation
in a nonlinear medium.

\begin{figure}[htbp]
  \centerline{
    \includegraphics[width=70mm]{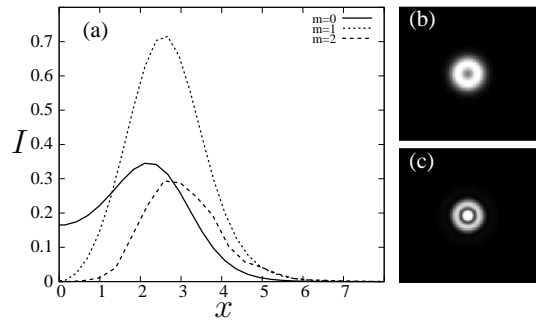}
  }
  \caption{
    An incoherent vortex soliton calculated using the modal theory and three
    modes with the topological charges $m=0,\, 1$ and $3$. (a) shows
    the profiles of the three components, (b) shows the total intensity
    of the soliton and (c) shows its cross-correlation
    $\Gamma(-\mathbf{r},\mathbf{r})$.}
  \label{fig2}
\end{figure}

Although the coherence density approach can be used to simulate
the propagation of partially incoherent light with arbitrary
precision, it is of little use when it comes to finding an
explanation for the results obtained from the simulations. More
physical insight into the problem can be obtained by using the
modal theory of incoherent solitons~\cite{MitchellPRL97}.
According to the modal theory, the incoherent solitons can be
regarded as an incoherent superposition of guided modes of the
waveguide induced by the total light intensity. Since the
incoherent vortices that we are dealing with induce circularly
symmetric waveguides, the guided modes we have to consider are
also circularly symmetric. To explain our numerical findings, we
construct a partially incoherent vortex soliton using circularly
symmetric beams with topological charges $m=0,\,1$ and $2$:
$E(\mathbf{r})=\sum_{m=0}^3
E_m(\mathbf{r})\exp(im\varphi)\exp(i\gamma_m(t))$. This can be
done using a standard relaxation technique~\cite{muell}. A more
precise modelling of incoherent solitons would require more modes.
Here, we restrict ourselves to {\em three modes} only, assuming
that for a partially incoherent vortex the $m=1$ component should
be dominant and that the next strongest components should be those
with topological charge $m^\prime=m\pm 1$, i.e. $m^\prime=0, \,
2$. Indeed, we find that the main features of incoherent vortex
solitons can be  explained qualitatively using only these three
modes.

\begin{figure}[htbp]
  \centerline{
    \includegraphics[width=50mm]{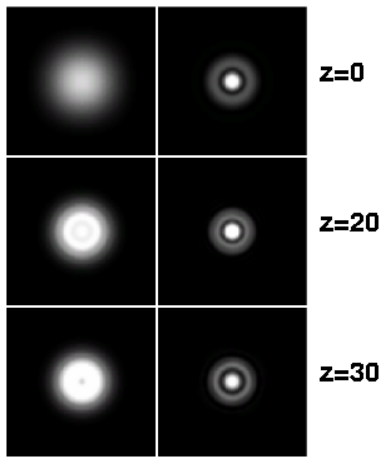}}
  \caption{
    The intensity (left column) and the cross-correlation (right column) of the
    far field. The effects of the nonlinearity on the intensity distribution
    can be clearly seen, whereas the cross-correlation maintains more or less
    the structure one would expect in the case of linear propagation.
  }
  \label{fig3}
\end{figure}

The relative intensity of the $m=0$ beam and the $m=2$ vortex as
compared to the $m=1$ vortex control the light coherence. However,
in order to assure that the total topological charge of the light
$m_{\rm tot}=\mbox{Im} \{\left<\int E^*(\mathbf{r}\times\nabla
E)\mbox{d}\mathbf{r}\right>\}\mathbf{e}_z/ \int
I\mbox{d}\mathbf{r}$ is equal to one, we have to chose the $m=0$
and $m=2$ components of equal intensity. In order to check whether
this simple approach yields the results that agree at least
qualitatively with the full numerical model, we calculate an
incoherent vortex soliton using the modal theory. The resulting
shape of the vortex components, the total intensity, and
cross-correlation $\Gamma(-\mathbf{r},\mathbf{r})$ are shown in
Fig.~\ref{fig2}. Comparing Fig.~\ref{fig1} and Fig.~\ref{fig2}, we
notice the presence of two similar features: the local minimum of
the intensity in the center of the beam, and the ring-like
structure of the cross-correlation. Hence, these two phenomena can
be explained by considering a simple modal representation of the
incoherent vortex consisting of only three modes with the
topological charges $m=0,\,1$ and $2$.

First, the local minimum in the center of the beam can be
explained by the fact that the waveguide induced by the $m=1$ and
$m=2$ components affects the $m=0$ mode in such a way, that it
also develops a local intensity minimum in its center, a fact well
known from the vortex-mode vector  solitons. Second, the ring-like
structure of the cross-correlation comes from the different radial
extent of the single components. As is known from the physics of
vortex-mode vector  solitons, the $m=0$ component has the smallest
radial extent, whereas the $m=1$ and $m=2$ components have larger
radii. Hence the cross-correlation given by
$\Gamma(-\mathbf{r},\mathbf{r})=\sum_{m,m^\prime=0}^3
\left<E^*_{m^\prime}(-\mathbf{r})E_m(\mathbf{r})\right>=
\sum_{m=0}^3 E^*_m(-\mathbf{r}) E_m(\mathbf{r})$, is dominated for
small $\mathbf{r}$ by the auto-correlated $m=0$ component, whereas
the $m=1$ component dominates for larger $\mathbf{r}$. For even
larger $\mathbf{r}$, the $m=2$ component can also come into play
which can eventually result in a second ring of auto-correlation.

Returning to the more precise, yet numerically more demanding
coherence density approach, we show in Fig.~\ref{fig3} the
situation in the far field. All parameters are identical to those
used in Fig.~\ref{fig1}. In the far field as well we observe a
ring-like structure of the cross-correlation function
$\Gamma(-\mathbf{f},\mathbf{f})$, where $\mathbf{f}$ stands for
the spatial coordinates in the far field. The intensity
distribution in the far field can also show a local minimum in the
center of the beam, contrary to what one would obtain if the
vortex was propagating through a linear medium
\cite{PalaciosPRL04}, and also in contrast to the result we would
obtain if we were propagating a light beam without topological
charge. This emphasizes the importance of the interaction between
the beam coherence function and the nonlinearity.

In conclusion, we have shown that the phase singularities in the
spatial coherence function found earlier in the linear optics can
survive the propagation through nonlinear media when the singular
beam creates an incoherent vortex soliton. Our results emphasize
the importance of the spatial coherence function in the studies of
the propagation of incoherent singular beams. Not only the phase
structure, but also the intensity distribution strongly depends on
the initial form of the coherence function of the light beam as it
enters a nonlinear medium.

Kristian Motzek is also at the Institute of Applied Physics,
Darmstadt University of Technology, Germany. The authors thank
Ming-feng Shih and Andrey Sukhorukov for helpful discussions.

\end{document}